\documentclass[12pt,a4paper]{article}
\usepackage{amsfonts}
\usepackage{amssymb}
\usepackage{amsmath}
\usepackage{latexsym}
\usepackage{graphicx}
\usepackage{hyperref}
\begin{document}

\begin{center}
{\Large \bf More on divergences in brane world models} \\

\vspace{4mm}

Mikhail N.~Smolyakov\\
\vspace{0.5cm} Skobeltsyn Institute of Nuclear Physics, Lomonosov
Moscow State University,
\\ 119991, Moscow, Russia\\
\end{center}

\begin{abstract}
In this paper a model in a space-time with compact extra dimension
is presented, describing five-dimensional fermion fields
interacting with an electromagnetic field localized on a brane.
This model can be considered as a toy model for examining possible
consequences of the localization of gauge fields on a brane. It is
shown that in the limit of infinite extra dimension, the
lowest-order amplitudes of some processes in the resulting
four-dimensional effective theory are divergent. Such a
"localization catastrophe" can be inherent to more realistic brane
world models with infinite extra dimensions.
\end{abstract}

In Ref.~\cite{Smolyakov}, an attempt was made to construct a model
with infinite extra dimension, describing spinor electrodynamics
localized on a domain wall. It was found that due to the existence
of nonlocalized fermion modes, though with large four-dimensional
masses, the regularized amplitude of the standard process of
quantum electrodynamics --- the light-by-light scattering --- is
divergent. This divergence appears to be a "physical" one; i.e.,
it cannot be removed by means of the standard renormalization
procedure. It simply reflects the fact that an infinite number of
fermions, as seen from the four-dimensional point of view, each
with the same coupling to the vector field, contribute to the
amplitude.

It is reasonable to suppose that such divergences can arise in
other models with infinite extra dimensions, where the zero mode
of the gauge field is supposed to be localized on a brane. To show
the origin of the problem in a simple way, we will consider a toy
model which possesses the same property as the one considered in
\cite{Smolyakov}. It will be shown that an initially local theory
with a gauge field localized on a brane can result in the same
effects as a theory with a nonlocal interaction between fermions
and a gauge field.

In the beginning, let us briefly discuss possible
field-theoretical mechanisms of field localization on a brane in a
space-time with one infinite extra dimension. The best-known
mechanism for localization of fermions is the Rubakov-Shaposhnikov
mechanism \cite{RS} of fermion localization on a domain wall (see
also its generalizations in \cite{DRT}). The resulting effective
theory contains a localized fermion (there can be more than one
localized mode for an appropriate choice of the parameters), whose
wave function in the extra dimension falls off exponentially, and
modes from the continuous spectrum. There is a nonzero mass gap
between the localized mode and the modes from the continuous
spectrum. As for the gauge fields, there are also certain
mechanisms of localization --- see, for example, \cite{KT}. To
illustrate the key idea of \cite{KT}, let us consider the action
for the Abelian gauge field of the form
\begin{equation}\label{actGloc}
S=-\int d^{4}xdz \frac{G(z)}{4}F_{MN}F^{MN},
\end{equation}
where $G(z)$ is some positive function such that $G(z)\to 0$ for
$z\to\pm\infty$. This function can either be defined by the fields
forming a domain wall or just be added to the model "by hand".
From the very beginning we can impose the gauge $A_{4}\equiv 0$.
From the equations of motion, following from (\ref{actGloc}), it
is clear that the solution for the lowest mode has the form
$A_{\mu}(x,z)=Ca_{\mu}(x)$, where $C$ is a constant and
$a_{\mu}(x)$ satisfies the Maxwell equations. By an appropriate
choice of the function $G(z)$, we can always make a nonzero mass
gap between the lowest mode and the other Kaluza-Klein modes
(particular realizations of such a scenario can be found in
\cite{Smolyakov,KT}). If the effective width of the function
$G(z)$ is rather small (i.e., it is much smaller than the other
characteristic sizes of the model), for an appropriate choice of
$C$, the action for the lowest mode can be approximated as
\begin{equation}\label{actGloc1}
S_{0}=-\int d^{4}xdz \frac{G(z)C^2}{4}f_{\mu\nu}f^{\mu\nu}\to
-\int d^{4}xdz \frac{\delta(z)}{4}f_{\mu\nu}f^{\mu\nu},
\end{equation}
where $f_{\mu\nu}=\partial_{\mu}a_{\nu}-\partial_{\nu}a_{\mu}$.
Note that the interaction of this massless mode with the fermions
living in the bulk is defined by the field
$A_{\mu}(x,z)=Ca_{\mu}(x)$ and looks like a nonlocal interaction,
though it originates from a consistent five-dimensional local
theory. Moreover, any consistent field-theoretical mechanism for
gauge field localization should lead to an effective theory
similar to the one described above: there should be a lowest
localized mode which is massless from the four-dimensional point
of view; its wave function in the extra dimension should be a
constant to prevent the universality of charge, which is crucial
in non-Abelian gauge theories (see a detailed discussion of this
problem in \cite{Rubakov}); and there should be a nonzero mass gap
between the lowest localized mode and the modes from the
continuous spectrum.

We will be interested in processes involving only the lowest
localized mode of the vector field (say, the photon) and all the
fermion modes. Of course, it is possible to consider the initial
five-dimensional theory and to calculate the amplitudes of the
corresponding processes in this theory. Meanwhile, such a
calculation can be rather complicated, and it lies beyond the
scope of the present paper. A much simpler way is to put our model
into a "box" of a finite size in the extra dimension to discretize
the modes from the continuous spectra, then calculate the
necessary amplitudes and take the limit of an infinite size of the
box. Such a procedure also allows one to see how the divergences,
which will be discussed below, arise when one passes from a
compact extra dimension to an infinite one.

The reasoning presented above suggests the following toy model in
a compact extra dimension. Let us take a model in a
five-dimensional space-time with the coordinates
$x^{M}=\{x^{\mu},z\}$, $M=0,1,2,3,4$, describing fermions
interacting with an Abelian gauge field. The compact extra
dimension with the coordinate $-L\le z\le L$ is supposed to form
an orbifold with the points $-z$ and $z$ identified. In what
follows, we will use the notation $x$ for the coordinates
$x^{\mu}$. The brane is supposed to be located at a fixed point of
the orbifold --- say, at the point $z=0$.

The total action of the model consists of two terms. The first
term has the form
\begin{equation}\label{SB}
S_{b}=\int
d^{4}x\left(-\frac{1}{4}f_{\mu\nu}f^{\mu\nu}+i\bar\psi_{b}\gamma^{\mu}\left(\partial_{\mu}-ie_{b}a_{\mu}\right)\psi_{b}-m\bar\psi_{b}\psi_{b}\right).
\end{equation}
It describes the fermion field $\psi_{b}$ living on the brane,
which is coupled to the vector field $a_{\mu}$ with the coupling
constant $e_{b}$. In our scenario, a particular mechanism of
localization is not taken into account. The second term includes
fermions living in the whole five-dimensional space-time. In order
to have a nonzero mass term for the zero Kaluza-Klein fermion mode
--- i.e., to have a nonzero mass gap between the localized fermion
and the lowest Kaluza-Klein mode in the effective four-dimensional
action --- we introduce two five-dimensional spinor fields (see,
for example, \cite{DRT,Macesanu}) possessing the orbifold symmetry
conditions
\begin{eqnarray}\label{sym1}
\Psi_{1}(x,-z)=\gamma^{5}\Psi_{1}(x,z),\\ \label{sym2}
\Psi_{2}(x,-z)=-\gamma^{5}\Psi_{2}(x,z).
\end{eqnarray}
Thus, the second term of the total action has the form
\begin{eqnarray}\label{sact}
S_{f}=\int d^{4}x\int\limits_{-L}^{L}dz
\left[i\bar\Psi_{1}\Gamma^{N}\left(\partial_{N}-ieA_{N}\right)\Psi_{1}
+i\bar\Psi_{2}\Gamma^{N}\left(\partial_{N}-ieA_{N}\right)\Psi_{2}-M\left(\bar\Psi_{1}\Psi_{2}+\bar\Psi_{2}\Psi_{1}\right)\right],
\end{eqnarray}
where $N=0,...,4$, $\Gamma^{\mu}=\gamma^{\mu}$,
$\Gamma^{4}=i\gamma^{5}$, $A_{4}\equiv 0$,
$A_{\mu}(x,z)=a_{\mu}(x)$. Formally, such a form of the
interaction cannot be discarded: the total action is invariant
under the four-dimensional $U(1)$ gauge group and under the
four-dimensional Lorentz transformations.

If the five-dimensional coupling constant $e=0$, then the
effective four-dimensional theory is just the standard
four-dimensional electrodynamics, and we do not expect any new
effects caused by the Kaluza-Klein modes. We will proceed with the
possibility $e\ne 0$, which is much more interesting. We think
that the most natural choice for the coupling constant $e_{b}$ is
$e_{b}=e$.

The four-dimensional effective action, coming from (\ref{sact}),
has the form (see the detailed derivation in Appendix~A)
\begin{eqnarray}\label{seff}
S_{f\_eff}=\int d^{4}x
\left.\Biggr[i\bar\psi\gamma^{\mu}\left(\partial_{\mu}-iea_{\mu}\right)\psi-M\bar\psi\psi\right.\\
\nonumber\left.+\sum\limits_{n=1}^{\infty}\sum\limits_{i=1}^{2}\left(
i\bar\psi_{i}^{n}\gamma^{\mu}\left(\partial_{\mu}-iea_{\mu}\right)\psi_{i}^{n}-
\mu_{n}\bar{\psi}_{i}^{n}\psi_{i}^{n}\right)\right]
\end{eqnarray}
with
\begin{equation}\label{eq8}
\mu_{n}=\sqrt{\frac{\pi^2n^2}{L^2}+M^2}.
\end{equation}
Now the total four-dimensional effective action, coming from
(\ref{SB}) and (\ref{sact}), has the form
\begin{eqnarray}\label{stot}
S_{eff}=\int d^{4}x
\left.\Biggl[-\frac{1}{4}f_{\mu\nu}f^{\mu\nu}+i\bar\psi_{b}\gamma^{\mu}\left(\partial_{\mu}-iea_{\mu}\right)\psi_{b}-m\bar\psi_{b}\psi_{b}\right.\\
\nonumber \left.
+i\bar\psi\gamma^{\mu}\left(\partial_{\mu}-iea_{\mu}\right)\psi-M\bar\psi\psi+\sum\limits_{n=1}^{\infty}\sum\limits_{i=1}^{2}\left(
i\bar\psi_{i}^{n}\gamma^{\mu}\left(\partial_{\mu}-iea_{\mu}\right)\psi_{i}^{n}-
\mu_{n}\bar{\psi}_{i}^{n}\psi_{i}^{n}\right)\right].
\end{eqnarray}
Of course, in a more general theory there can be massive
Kaluza-Klein modes of the vector field, which can be described by
the extra term
\begin{equation}\label{vect-extra}
S_{extra}=\int d^{4}x\int\limits_{-L}^{L}dz
\left(-\frac{1}{4}{\hat F_{MN}}{\hat F}^{MN}+\frac{1}{2}{\hat
M}^2{\hat A}^{M}{\hat A}_{M}\right),
\end{equation}
and the corresponding interactions of the field ${\hat A}_{M}$
with the fermion fields in the action of our toy model (it is
clear that the spectrum of Kaluza-Klein modes of the field ${\hat
A}_{M}$ does not contain a massless four-dimensional mode), but
these massive Kaluza-Klein modes will be irrelevant for our
analysis, so we just drop them from our toy model and the
corresponding four-dimensional effective action. This is a
reasonable assumption --- apart from the massive vector
Kaluza-Klein modes, the discretized effective four-dimensional
action of a more complicated model presented in \cite{Smolyakov}
is very similar to (\ref{stot}). Thus, though the interaction
between the five-dimensional fermions and the gauge field in
(\ref{sact}) looks rather strange because of its nonlocality, such
an action indeed can result from a consistent five-dimensional
local theory, if we retain only the lowest Kaluza-Klein mode of
the vector field. In other words, the action described by
(\ref{SB}) and (\ref{sact}) can be considered as a toy model,
where a particular mechanism of localization is not taken into
account, but the localized theory (described by $\psi_{b}$ and
$a_{\mu}$) "remembers" the fact that it originates from a more
general five-dimensional theory through the nonlocal interaction
described by (\ref{sact}) (the fields $\Psi_{1}$ and $\Psi_{2}$
can be considered as the fields describing the nonlocalized
fermion part of the initial five-dimensional theory). Note that,
in principle, the toy model described by (\ref{SB}) and
(\ref{sact}) can be constructed using only the general
requirements for any field-theoretical mechanism for gauge field
localization on the brane, which were mentioned above.

We also suppose that $M\gg m$; i.e., the mass gap between the
brane localized theory and the Kaluza-Klein modes is very large.
The parameter $M$ can be considered as the energy scale at which
five-dimensional effects may come into play.

\begin{figure}[ht]
\center{\includegraphics[width=0.99\linewidth]{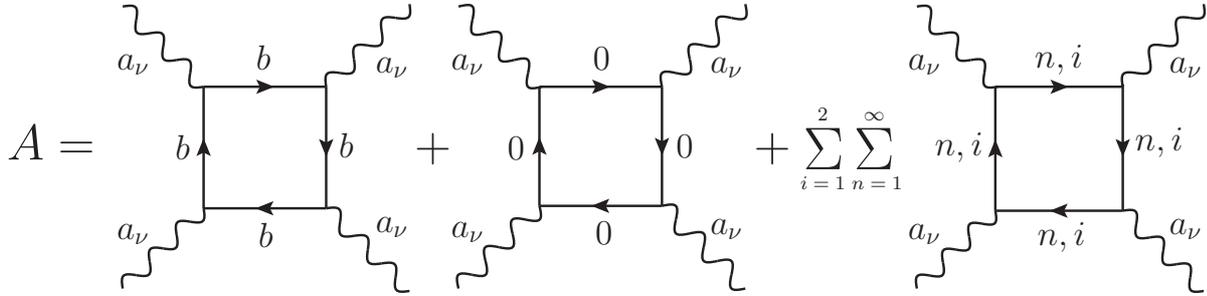}}
\caption{{\footnotesize Diagrams corresponding to the
$\gamma\gamma$ scattering amplitude. The number marking each
fermion line corresponds to the number of the massive fermion
Kaluza-Klein mode which is represented by the line. The letter $b$
stands for a brane-localized fermion.}} \label{fig1}
\end{figure}

Now we are ready to consider particular effects following from
action (\ref{stot}). As in \cite{Smolyakov}, we will be interested
in $\gamma\gamma\to\gamma\gamma$ scattering, where $\gamma$ stands
for the particle corresponding to the vector field $a_{\mu}$ (say,
the photon). According to action (\ref{stot}), the corresponding
amplitude in the lowest order in the coupling constant is
schematically represented in Figure~\ref{fig1}.
For $\omega\ll m$, where $\omega$ is the energy of the photon in
the c.m. frame, the regularized amplitude in the leading order in
$\frac{\omega}{m}$  is given by \cite{AB}
\begin{equation}\label{amplsum}
A=\left(\frac{e^{4}\omega^4}{m^4}+\frac{e^{4}\omega^4}{M^4}+2\sum\limits_{n=1}^{\infty}\frac{e^{4}\omega^4}{\mu_{n}^4}\right)
F(\theta),
\end{equation}
where the function $F(\theta)$ depends on the scattering angle
$\theta$ and the polarizations of the photons. We can estimate the
series in $A$ using the standard Maclaurin-Cauchy integral test
for convergence:
\begin{eqnarray}\label{estim}
\frac{L}{4M^{3}}-\frac{L\arctan\left(\frac{\pi}{ML}\right)}{2\pi
M^{3}}-\frac{L^{2}}{2M^{2}(\pi^2+M^{2}L^{2})}=\int_{1}^{\infty}dx\frac{L^4}{\left(\pi^2x^2+M^2L^2\right)^2}\\
\nonumber
<\sum\limits_{n=1}^{\infty}\frac{1}{\mu_{n}^4}=\sum\limits_{n=1}^{\infty}\frac{L^4}{\left(\pi^2n^2+M^2L^2\right)^2}<
\int_{0}^{\infty}dx\frac{L^4}{\left(\pi^2x^2+M^2L^2\right)^2}=\frac{L}{4M^{3}}.
\end{eqnarray}
Thus, amplitude (\ref{amplsum}) can be estimated as
\begin{eqnarray}\label{ampl1}
A>e^{4}\omega^4F(\theta)\left
(\frac{1}{m^4}+\frac{1}{M^4}+\frac{L}{2M^{3}}-\frac{L\arctan\left(\frac{\pi}{ML}\right)}{\pi
M^{3}}-\frac{L^{2}}{M^{2}(\pi^2+M^{2}L^{2})}\right).
\end{eqnarray}
One can see that in the limit $L\to\infty$, i.e., when we pass to
an infinite extra dimension, the amplitude $A\to\infty$. Note that
though the mass gaps between the Kaluza-Klein modes tend to zero
in this limit, the mass gap $\Delta m=M-m$ between the
brane-localized fermion and the lowest Kaluza-Klein mode remains
intact, and it can be very large. A divergence of exactly the same
type arises in the model discussed in \cite{Smolyakov}. This
"localization catastrophe" is the consequence of the fact that,
though the vector field is localized on the brane, it interacts
with the five-dimensional fermions everywhere in the bulk. Such a
nonlocality of the interaction leads to a pathology, and it can
arise in more realistic scenarios of gauge field localization on
the brane.

It should be noted that the use of a finite cutoff scale does not
solve the problem. Indeed, let us take a cutoff scale $\tilde M$
such that we do not take into account Kaluza-Klein modes with the
masses $\mu_{n}>\tilde M$. The number of the heaviest Kaluza-Klein
mode, which contributes to the amplitude, can be defined as
$N=\lfloor\frac{L\sqrt{{\tilde M}^2-M^2}}{\pi}\rfloor$, where
$\lfloor\cdot\rfloor$ stands for the floor function. Thus,
\begin{equation}\label{eq14}
\sum\limits_{n=1}^{N}\frac{1}{\mu_{n}^4}>\sum\limits_{n=1}^{N}\frac{1}{{\tilde
M}^4}=N\frac{1}{{\tilde M}^4}>\left(\frac{L\sqrt{{\tilde
M}^2-M^2}}{\pi}-1\right)\frac{1}{{\tilde M}^4},
\end{equation}
which tends to infinity in the limit $L\to\infty$ for $\tilde
M>M$. Indeed, in the limit $L\to\infty$ the masses of the modes
$\mu_{n}\to M$. Thus, the larger $L$ becomes, the greater the
number of modes that appear below the cutoff scale $\tilde M$,
each with the same coupling constant to the gauge field which does
not depend on $L$, leading to an amplitude which increases with
increasing $L$ (see also Appendix~B for additional discussion of
this issue). Formally, this problem can be solved by considering
the cutoff scale such that some constant finite number of the
modes $N_{f}$ is below the cutoff scale for any $L$. But in the
limit $L\to\infty$, the corresponding cutoff scale $\tilde
M_{f}\to M$, which is obviously unphysical. Moreover, in the limit
$L\to\infty$, the discrete spectrum turns into the continuous one
without isolated modes, and a fixed $N_{f}$ in this case means
that the contribution from the continuous spectrum is of the
measure zero --- we just remove the continuous spectrum from the
theory and do not estimate its contribution. So, here and below,
we consider $\tilde M$ such that it does not depend on $L$, and
$\tilde M>M$.

\begin{figure}[ht]
\center{\includegraphics[width=0.99\linewidth]{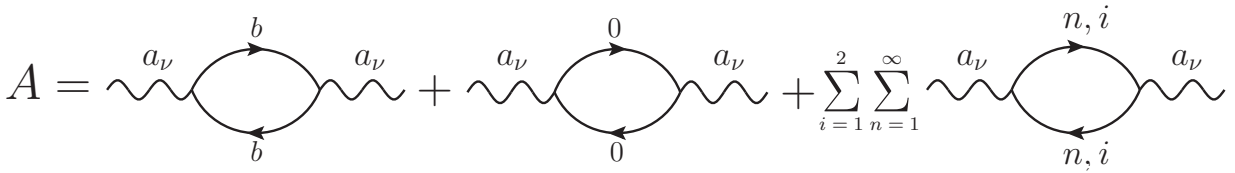}}
\caption{{\footnotesize Diagrams which contribute to the
polarization operator of the field $a_{\nu}$ in the leading order.
As in Figure~\ref{fig1}, the number marking each fermion line
corresponds to the number of the massive fermion Kaluza-Klein mode
which is represented by the line, and the letter $b$ stands for a
brane-localized fermion.}} \label{fig2}
\end{figure}

One can expect that effects analogous to the one which was
demonstrated above can arise in other processes, such that the
corresponding Feynaman diagrams contain the field $a_{\mu}$ in
external lines and the fermion fields in internal lines. For
example, let us take the polarization operator of the field
$a_{\mu}$. The corresponding diagrams are presented in
Figure~\ref{fig2}.

The polarization operator has the form
\begin{equation}
\Pi_{\mu\nu}(q)=(q_{\mu}q_{\nu}-\eta_{\mu\nu}q^{2})\left(\Pi(q^2,m_b)+\Pi(q^2,\mu_0)+2\sum\limits_{n=1}^{\infty}\Pi(q^2,\mu_n)\right).
\end{equation}
It is well known that for $q^2<m_{b}^2$ the function
$\Pi(q^2,m_{b})$ can be expanded as \cite{Schweber}
\begin{equation}
\Pi(q^2,m_b)=\Pi(0,m_b)+\frac{\partial \Pi(q^2,m_b)}{\partial
q^2}\biggl|_{q^2=0}q^{2}+\cdot\cdot\cdot,
\end{equation}
where $\frac{\partial \Pi(q^2,m_b)}{\partial
q^2}\bigl|_{q^2=0}=\frac{16i\pi^2}{60 m_{b}^2}$, and $\Pi(0,m_b)$
is logarithmically divergent. An analogous expansion can be made
for the contributions of the other Kaluza-Klein fermions. Thus, in
the leading order in $q^2/(\textrm{mass of the mode})^2$, the
regularized contribution to the polarization operator is
proportional to
\begin{equation}\label{pol}
q^{2}\left(\frac{1}{m_{b}^2}+\frac{1}{\mu_{0}^2}+2\sum\limits_{n=1}^{\infty}\frac{1}{\mu_{n}^2}\right).
\end{equation}
In the limit $L\to\infty$, the sum in (\ref{pol}) also diverges.
To show it, let us again introduce a cutoff scale $\tilde M>M$. In
this case,
\begin{equation}\label{pol1}
\sum\limits_{n=1}^{\infty}\frac{1}{\mu_{n}^2}>\sum\limits_{n=1}^{\lfloor\frac{L\sqrt{{\tilde
M}^2-M^2}}{\pi}\rfloor}\frac{1}{\mu_{n}^2}>\sum\limits_{n=1}^{\lfloor\frac{L\sqrt{{\tilde
M}^2-M^2}}{\pi}\rfloor}\frac{1}{{\tilde
M}^2}=\frac{\lfloor\frac{L\sqrt{{\tilde
M}^2-M^2}}{\pi}\rfloor}{{\tilde M}^2}>\left(\frac{L\sqrt{{\tilde
M}^2-M^2}}{\pi}-1\right)\frac{1}{{\tilde M}^2}.
\end{equation}
One can see that (\ref{pol1}) indeed is divergent in the limit
$L\to\infty$, so the polarization operator diverges for $q^{2}\ne
0$.

Just for a comparison, let us consider a rather different scenario
--- where the vector field can freely propagate in the bulk. In
this case, the action for the vector field can be chosen to be
\begin{equation}\label{SAbulk}
S_{vect}=-\int d^{4}xdz\frac{\xi^{2}}{4}F_{MN}F^{MN},
\end{equation}
instead of the corresponding brane-localized term in (\ref{SB}).
Here the constant $\xi$ with the dimension
$\sqrt{\textrm{[mass]}}$ is introduced for convenience --- the
field $A_{M}$ in this case has the standard dimension
$\textrm{[mass]}$. From the very beginning we can impose the gauge
$A_{4}\equiv 0$. After imposing this gauge, we are left with the
residual gauge transformations which are responsible for isolating
the physical degrees of freedom of the massless four-dimensional
vector field. We will be interested in the lowest mode of the
vector field, so we take the ansatz
$A_{\mu}(x,z)=\frac{1}{\xi\sqrt{2L}}a_{\mu}(x)$. Substituting it
into action (\ref{SAbulk}), integrating over the coordinate of the
extra dimension, and taking into account (\ref{seff}) and
(\ref{SB}), we obtain
\begin{eqnarray}\label{stotbulk}
S=\int
d^{4}x\left.\Biggl(-\frac{1}{4}f_{\mu\nu}f^{\mu\nu}+i\bar\psi_{b}\gamma^{\mu}\left(\partial_{\mu}-ie_{4}a_{\mu}\right)\psi_{b}-m\bar\psi_{b}\psi_{b}\right.\\
\nonumber \left.
+i\bar\psi\gamma^{\mu}\left(\partial_{\mu}-ie_{4}a_{\mu}\right)\psi-M\bar\psi\psi+\sum\limits_{n=1}^{\infty}\sum\limits_{i=1}^{2}\left(
i\bar\psi_{i}^{n}\gamma^{\mu}\left(\partial_{\mu}-ie_{4}a_{\mu}\right)\psi_{i}^{n}-
\mu_{n}\bar{\psi}_{i}^{n}\psi_{i}^{n}\right) \right),
\end{eqnarray}
where $e_{4}=\frac{e}{\xi\sqrt{2L}}$.

Now let us calculate the amplitude corresponding to the process
presented in Figure~\ref{fig1}. It has the form
\begin{equation}\label{amplsum1}
A=\left(\frac{e_{4}^{4}\omega^4}{m^4}+\frac{e_{4}^{4}\omega^4}{M^4}+2\sum\limits_{n=1}^{\infty}\frac{e_{4}^{4}\omega^4}{\mu_{n}^4}\right)
F(\theta)=\frac{1}{4L^2\xi^{4}}\left(\frac{e^{4}\omega^4}{m^4}+\frac{e^{4}\omega^4}{M^4}+2\sum\limits_{n=1}^{\infty}\frac{e^{4}\omega^4}{\mu_{n}^4}\right)
F(\theta).
\end{equation}
Using (\ref{estim}) we can estimate the amplitude as
\begin{eqnarray}
A<\frac{e^{4}\omega^4F(\theta)}{4\xi^{4}}\left
(\frac{1}{L^{2}}\left(\frac{1}{m^4}+\frac{1}{M^4}\right)+\frac{1}{2M^{3}L}\right).
\end{eqnarray}
We see that now in the limit $L\to\infty$, the amplitude $A\to 0$.
From the four-dimensional point of view, the only difference
between this case and the previous one with the brane-localized
gauge field is in the value of the coupling constant ($e$ versus
$e/(\xi\sqrt{2L})$). From the physical point of view the
difference is obvious --- the second case corresponds to a local
theory. Thus, in the limit $L\to\infty$, the zero mode of the
gauge field $A_{\mu}$ ceases to be a four-dimensional particle ---
there is no mass gap between the zero mode and the next
Kaluza-Klein mode of the vector field, and between the other
Kaluza-Klein modes. Now all the Kaluza-Klein modes compose a
single five-dimensional particle. In the resulting
five-dimensional (though non-renormalizable in the usual sense)
theory, the corresponding amplitude, of course, is nonzero.

Let us look at this case from another point of view --- again by
introducing the cutoff scale $\tilde M$. For the amplitude, we
have
\begin{equation}\label{bulkcut-off}
\sum\limits_{n=1}^{N}\frac{e_{4}^{4}}{\mu_{n}^4}<N\frac{e_{4}^{4}}{{
M}^4}<\left(\frac{L\sqrt{{\tilde
M}^2-M^2}}{\pi}-1\right)\frac{e^{4}}{4\xi^{4}L^{2}}\frac{1}{{M}^4},
\end{equation}
which tends to zero when $L\to\infty$. Again, the larger $L$
becomes, the greater the number of modes that appear below the
cutoff scale $\tilde M$, but now the effective coupling constant
$e_{4}$ of each mode to the gauge field depends on $L$, leading to
the amplitude which decreases with increasing $L$.

Finally, let us consider the last example --- the model of
quasilocalization of a gauge field proposed in \cite{QL} (a
particular realization of this mechanism can be found in
\cite{QL1}). The action for the gauge field takes the form (in our
notations)
\begin{equation}\label{quasilocact}
S_{vect}=-\int
d^{4}xdz\left(\frac{\xi^{2}}{4}F_{MN}F^{MN}+\delta(z)\frac{1}{4}F_{\mu\nu}F^{\mu\nu}\right).
\end{equation}
As in the previous cases, let us put this system into a box in the
extra dimension; i.e., let us suppose that the extra dimension
forms an orbifold with the size $2L$. Again, we can impose the
gauge $A_{4}\equiv 0$ and expand the vector field into
Kaluza-Klein modes. The lowest mode (which we will be interested
in) of the vector field has a constant wave function in the extra
dimension (which follows from the corresponding equations of
motion for the vector field), so we can take the ansatz
$A_{\mu}(x,z)=\frac{1}{\sqrt{2\xi^{2} L+1}}a_{\mu}(x)$. After
substituting it into (\ref{quasilocact}), integrating over the
coordinate of the extra dimension, and taking into account
(\ref{sact}), we obtain (\ref{stotbulk}), but now with
$e_{4}=\frac{e}{\sqrt{2\xi^{2} L+1}}$. Instead of
(\ref{bulkcut-off}), we get
\begin{equation}\label{bulkcut-off1}
\sum\limits_{n=1}^{N}\frac{e_{4}^{4}}{\mu_{n}^4}<N\frac{e_{4}^{4}}{{
M}^4}<\left(\frac{L\sqrt{{\tilde
M}^2-M^2}}{\pi}-1\right)\frac{e^{4}}{\left(2\xi^{2}
L+1\right)^{2}}\frac{1}{{M}^4},
\end{equation}
which also tends to zero when $L\to\infty$. As in the previous
case with the gauge field living in the bulk, in the limit
$L\to\infty$, the zero mode of the gauge field $A_{\mu}$ ceases to
be a four-dimensional particle --- there is no mass gap between
the zero mode and the next Kaluza-Klein mode of the vector field,
and between the other Kaluza-Klein modes (this happens because of
the existence of the term $\frac{\xi^{2}}{4}F_{MN}F^{MN}$ in
(\ref{quasilocact})). Again, we cannot isolate the purely
four-dimensional massless photon (a localized zero mode is absent
in such a scenario --- the photon behaves as a four-dimensional
particle at small distances, whereas at large distances it
possesses a five-dimensional behavior; see \cite{QL} for details),
contrary to the case of the localized gauge field described by
equations (\ref{SB}) and (\ref{sact}), where the mass gap between
the lowest localized mode and the rest Kaluza-Klein tower (which
can be taken into account by adding (\ref{vect-extra}) to
(\ref{SB}) and (\ref{sact})) remains nonzero even in the limit
$L\to\infty$, in full analogy with \cite{Smolyakov}.

An interesting observation is that (\ref{quasilocact}) turns into
the boson part of (\ref{SB}) if one takes $\xi=0$. But the choice
$\xi=0$ means that there is a strong coupling outside the brane in
our toy model (this also follows from the five-dimensional action
in (\ref{actGloc}) --- the "coupling constant"
$\frac{1}{\sqrt{G(z)}}$ grows with $z$). It is obvious that this
strong coupling is directly connected to the divergences discussed
above.

In conclusion, we have considered the simplest case of an Abelian
gauge field. It was shown that the localization of such a gauge
field on a brane can result in an effective action (as an example,
one can consider the model discussed in \cite{Smolyakov})
analogous to those coming from multidimensional theories
containing nonlocal interactions from the very beginning. Since
one may expect pathologies in theories with nonlocal interactions,
analogous pathologies can arise in theories with brane-localized
gauge fields in more general cases. One should take this effect
into account when considering brane world models with infinite
extra dimensions.

\section*{Acknowledgements}
The author is grateful to E.~Boos, M.~Iofa, D.~Kirpichnikov, and
I.~Volobuev for discussions and to the unknown referee for useful
comments. The work was supported by grant of the Russian Ministry
of Education and Science (agreement No.~8412), grants
NS-3920.2012.2 and MK-3977.2011.2 of the President of Russian
Federation, and RFBR grants \mbox{12-02-93108-CNRSL-a} and
10-02-00525-a. The
\href{http://jaxodraw.sourceforge.net/}{JaxoDraw} program package
\cite{JD} was used to draw the Feynman diagrams presented in
Figs.~1--3.

\section*{Appendix A}
According to the orbifold symmetry conditions (\ref{sym1}) and
(\ref{sym2}), the five-dimensional fermion fields $\Psi_{1}$ and
$\Psi_{2}$ can be decomposed into Kaluza-Klein modes (see
\cite{Macesanu}) as
\begin{eqnarray}\label{subst1}
\Psi_{1}(x,z)=\frac{1}{\sqrt{2L}}\psi_{L}(x)+\frac{1}{\sqrt{L}}\sum\limits_{n=1}^{\infty}\left(\cos\left(\frac{\pi
n}{L}z\right)\psi^{n}_{L}(x)-\sin\left(\frac{\pi
n}{L}z\right)\hat\psi^{n}_{R}(x)\right),\\ \label{subst2}
\Psi_{2}(x,z)=\frac{1}{\sqrt{2L}}\psi_{R}(x)+\frac{1}{\sqrt{L}}\sum\limits_{n=1}^{\infty}\left(\cos\left(\frac{\pi
n}{L}z\right)\psi^{n}_{R}(x)+\sin\left(\frac{\pi
n}{L}z\right)\hat\psi^{n}_{L}(x)\right),
\end{eqnarray}
where $\psi_{L}(x)=\gamma^{5}\psi_{L}(x)$,
$\psi_{R}(x)=-\gamma^{5}\psi_{R}(x)$,
$\psi^{n}_{L}(x)=\gamma^{5}\psi^{n}_{L}(x)$,
$\psi^{n}_{R}(x)=-\gamma^{5}\psi^{n}_{R}(x)$,
$\hat\psi^{n}_{L}(x)=\gamma^{5}\hat\psi^{n}_{L}(x)$,
$\hat\psi^{n}_{R}(x)=-\gamma^{5}\hat\psi^{n}_{R}(x)$. Substituting
(\ref{subst1}) and (\ref{subst2}) into (\ref{sact}) and
integrating over the coordinate $z$ of the extra dimension, we
arrive at
\begin{eqnarray}\label{act4stand}
S_{f\_eff}=\int d^{4}x
\left[i\bar\psi\gamma^{\mu}\left(\partial_{\mu}-iea_{\mu}\right)\psi-M\bar\psi\psi+\sum\limits_{n=1}^{\infty}\left.\Bigl(
i\bar\psi^{n}\gamma^{\mu}\left(\partial_{\mu}-iea_{\mu}\right)\psi^{n}\right.\right.\\
\nonumber \left.\left.+i\bar{\hat\psi}^{n}\gamma^{\mu}\left(
\partial_{\mu}-iea_{\mu}\right)\hat\psi^{n}-\frac{\pi n}{L}(\bar{\hat\psi}^{n}\psi^{n}+\bar\psi^{n}\hat\psi^{n})-
M(\bar\psi^{n}\psi^{n}-\bar{\hat\psi}^{n}\hat\psi^{n})\right)\right.\Biggr]
\end{eqnarray}
with
\begin{eqnarray}
\psi(x)=\psi_{L}(x)+\psi_{R}(x),\\
\psi^n(x)=\psi_{L}^n(x)+\psi_{R}^n(x),\\
\hat\psi^n(x)=\hat\psi_{L}^n(x)+\hat\psi_{R}^n(x).
\end{eqnarray}
We see that the mass matrix is nondiagonal. In order to bring it
into a diagonal form, we use the transformations
\begin{eqnarray}
\psi^{n}(x)&=&\psi_{1}^{n}(x)\cos(\theta_{n})+\psi_{2}^{n}(x)\sin(\theta_{n}),\\
\hat\psi^{n}(x)&=&\psi_{1}^{n}(x)\sin(\theta_{n})-\psi_{2}^{n}(x)\cos(\theta_{n})
\end{eqnarray}
with
\begin{equation}
\tan(2\theta_{n})=\frac{\pi n}{ML}
\end{equation}
and obtain
\begin{eqnarray}\label{act4unc}
S_{f\_eff}=\int d^{4}x
\left[i\bar\psi\gamma^{\mu}\left(\partial_{\mu}-iea_{\mu}\right)\psi-M\bar\psi\psi+\sum\limits_{n=1}^{\infty}\left(
i\bar\psi_{1}^{n}\gamma^{\mu}\left(\partial_{\mu}-iea_{\mu}\right)\psi_{1}^{n}\right.\right.\\
\nonumber
\left.\left.+i\bar\psi_{2}^{n}\gamma^{\mu}\left(\partial_{\mu}-iea_{\mu}\right)\psi_{2}^{n}-
\mu_{n}(\bar{\psi}_{1}^{n}\psi_{1}^{n}-\bar{\psi}_{2}^{n}\psi_{2}^{n})\right)\right.\Biggr],
\end{eqnarray}
where $\mu_{n}=\sqrt{\frac{\pi^2n^2}{L^2}+M^2}$. We see that the
mass terms of the fields $\psi_{2}^{n}$ have unconventional signs.
But with the help of the standard redefinition $\psi_{2}^{n}\to
\gamma^{5}\psi_{2}^{n}$, we can bring the action of
(\ref{act4unc}) into the standard form of (\ref{seff}).

It should be noted that one can find a similar doubling of the
number of four-dimensional effective degrees of freedom in the
case of one five-dimensional massive fermion in a space-time with
one compact extra dimension, where the extra dimension does not
possess the orbifold symmetry. Thus, the number of effective
four-dimensional degrees of freedom in the case of two
five-dimensional fermions in a space-time with a compact extra
dimension forming the $S^1/Z_2$ orbifold --- and in the case of a
single five-dimensional fermion in a space-time with a compact
extra dimension without the orbifold symmetry --- is the same;
i.e., the existence of the orbifold symmetry reduces almost by
half the number of Kaluza-Klein modes coming from each
five-dimensional fermion.

\section*{Appendix B}
Suppose the size of the extra dimension is rather large: $ML\gg
1$. In this case, the mass gap between two successive modes takes
the form
\begin{equation}
\Delta\mu_{n}=\mu_{n+1}-\mu_{n}\approx\frac{\mu_{n+1}^2-\mu_{n}^2}{2\mu_{n}}=\frac{\pi^{2}(2n+1)}{2L^{2}\mu_{n}}
=\frac{\pi}{L\mu_{n}}\left(\sqrt{\mu_{n}^{2}-M^{2}}+\frac{\pi}{2L}\right),
\end{equation}
where we have used equation (\ref{eq8}). Now we can rewrite the
first term of (\ref{eq14}) as
\begin{equation}\label{sumtoint}
\sum\limits_{n=1}^{N}\frac{1}{\mu_{n}^4}=\sum\limits_{n=1}^{N}\frac{\Delta\mu_{n}}{\Delta\mu_{n}}\frac{1}{\mu_{n}^4}\approx
\sum\limits_{n=1}^{N}\frac{L\Delta\mu_{n}}{\pi\mu_{n}^3\left(\sqrt{\mu_{n}^{2}-M^{2}}+\frac{\pi}{2L}\right)}\approx\frac{L}{\pi}\int_{M}^{\tilde
M}\frac{d\mu}{\mu^3\left(\sqrt{\mu^{2}-M^{2}}+\frac{\pi}{2L}\right)}.
\end{equation}
For any fixed cutoff scale $\tilde M>M$, the integral in
(\ref{sumtoint}) is finite. The larger $L$ becomes, the better the
integral term in (\ref{sumtoint}) approximates the initial sum. In
the limit $L\to\infty$, the integral itself remains finite, but
due to the factor $L$ in front of it, the amplitude tends to
infinity.

\begin{figure}[ht]
\center{\includegraphics[width=0.25\linewidth]{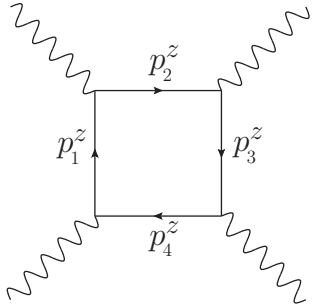}}
\caption{{\footnotesize Diagram corresponding to the
$\gamma\gamma$ scattering amplitude with five-dimensional
particles in internal lines. $p^{z}_{i}$, where $i=1,...,4$; stand
for the momenta of the five-dimensional fermions in the extra
dimension.}} \label{fig3}
\end{figure}

Integrals analogous to the one in (\ref{sumtoint}) are expected to
arise in models with infinite extra dimension. But the origin of
the factor $L$ in front of the integral in (\ref{sumtoint}) is not
obvious. Below, we will show schematically how it arises in a
model with initially infinite extra dimension. Indeed, the
existence of a continuous spectrum of excitations in effective
four-dimensional theory implies that there is a five-dimensional
particle which can move along the extra dimension. It possesses a
momentum $p^{z}$ along the extra coordinate $z$, varying in the
limits $M\le p^{z}\le \tilde M$, where the upper limit is defined
by our choice of the cutoff scale (of course, in the general case,
the particle can move in the opposite direction; i.e., we should
also take $-M\ge p^{z}\ge-\tilde M$, but it is irrelevant for our
analysis). Now let us consider the $\gamma\gamma\to\gamma\gamma$
scattering process, but with the five-dimensional particle in the
internal lines instead of a set of four-dimensional particles; see
Figure~\ref{fig3}.
The corresponding matrix element contains a term of the form
\begin{eqnarray}\label{5Damplitude}
\int dp^{z}_{1}\int dp^{z}_{2}\int dp^{z}_{3}\int
dp^{z}_{4}\,\delta(p^{z}_{1}-p^{z}_{2})\delta(p^{z}_{2}-p^{z}_{3})\delta(p^{z}_{3}-p^{z}_{4})
\delta(p^{z}_{4}-p^{z}_{1})F(p^{z}_{1},p^{z}_{2},p^{z}_{3},p^{z}_{4}),
\end{eqnarray}
where $F(p^{z}_{1},p^{z}_{2},p^{z}_{3},p^{z}_{4})$ is a function
depending on the momenta of fermions in internal lines in the
extra dimension [for simplicity, the four-dimensional momenta of
particles in external and internal lines of the diagram in
Figure~\ref{fig3}, together with the corresponding integrals, are
included in the definition of the function
$F(p^{z}_{1},p^{z}_{2},p^{z}_{3},p^{z}_{4})$]. Note that localized
photons do not carry momenta in the extra dimension; i.e., we
cannot say that a localized four-dimensional photon is just a
five-dimensional particle with $p^{z}=0$ --- that is why $\delta$
functions in (\ref{5Damplitude}) contain momenta in the extra
dimension only of the five-dimensional fermions. The integrals in
(\ref{5Damplitude}) can be easily evaluated, and we get
\begin{eqnarray}\label{5Damplitude1}
\int_{M}^{\tilde M}
dp^{z}_{1}\,\delta(p^{z}_{1}-p^{z}_{1})F(p^{z}_{1},p^{z}_{1},p^{z}_{1},p^{z}_{1})=\delta(0)\int_{M}^{\tilde
M} dp^{z}_{1}\,F(p^{z}_{1},p^{z}_{1},p^{z}_{1},p^{z}_{1}).
\end{eqnarray}
Though the integral on the RHS of (\ref{5Damplitude1}) can be
finite (at least after a four-dimensional regularization), the
value of (\ref{5Damplitude1}) is infinite because of the factor
$\delta(0)$. But
$\delta(0)=\frac{1}{2\pi}\int_{-\infty}^{\infty}dz=\frac{L}{\pi}$
(where $L\to\infty$), which is nothing but the size of the extra
dimension.\footnote{Analogous factors $\delta(0)$ arise when one
calculates disconnected diagrams in the standard four-dimensional
quantum field theory. Such diagrams make a contribution to the
vacuum energy; this contribution is proportional to
$\delta^{(3)}(0)$ (i.e., to the three-dimensional volume), and
thus it is infinite. In our case the situation is similar --- the
factor $\delta(0)$ in (\ref{5Damplitude1}) stands for a
contribution of the {\it whole} extra dimension to the
four-dimensional amplitude.} Thus, we have demonstrated how the
factor $L$ arises in model with initially infinite extra
dimension.

\end{document}